\documentclass[showpacs,preprintnumbers,amsmath,amssymb]{revtex4}
\usepackage{graphicx}
\usepackage{bm}
\usepackage{dcolumn}

\def\prl#1#2#3{{ Phys.   Rev.   Lett.  } {\bf #1}, #2 (#3)}
\def\pla#1#2#3{Phys.   Lett.   A {\bf #1}, #2 (#3)}

\def\pre#1#2#3{Phys.   Rev.   E {\bf #1}, #2 (#3)}

\def\pra#1#2#3{Phys.   Rev.   A {\bf #1}, #2 (#3)}

\def\jpa#1#2#3{J.   Phys.   A {\bf #1}, #2 (#3)}
\def\jsp#1#2#3{J.   Stat.   Phys.   {\bf #1}, #2 (#3)}

\def\rmp#1#2#3{Rev.   Mod.   Phys.   {\bf #1}, #2 (#3)}

\def\sci#1#2#3{Science {\bf#1},#2 (#3)}
\def\physa#1#2#3{Physica A {\bf#1},#2 (#3)}

\def\nl#1#2#3{Neture(London) {\bf#1},#2 (#3)}

\def\epl#1#2#3{Europhys. Lett. {\bf #1}, #2 (#3)}

\def\epjd#1#2#3{Eur.  Phys. J.  D {\bf #1}, #2 (#3)}
\def\pra#1#2#3{Phys.   Rev.   A {\bf #1}, #2 (#3)}

\def\svp#1#2#3{Sov. Phys.-JETP {\bf #1}, #2 (#3)}

\def\epl#1#2#3{{Europhys. Lett.} {\bf #1}, #2 (#3)}

\def\pre#1#2#3{Phys.   Rev.   E {\bf #1}, #2 (#3)}
\def\pknaw#1#2#3{Prok. Kon. N. Akad. Wet {\bf #1}, #2 (#3)}

\def\noi{\noindent}
\def\bc{\begin{center}}
\def\ec{\end{center}}
 \newcommand{\bea}{\begin{equation}}
 \newcommand{\eea}{\end{equation}\noi}
 \newcommand{\ber}{\begin{eqnarray}}
 \newcommand{\eer}{\end{eqnarray}\noi}
\begin{document}
\title{Finite size effect on Bose-Einstein condensation}
\author{Shyamal Biswas}\email{tpsb@iacs.res.in}
\affiliation{Department of Theoretical Physics,
Indian Association for the Cultivation of Science \\
Jadavpur,Kolkata-700032, India}
\date{\today}
\begin{abstract} 
               We show various aspects of finite size effects on Bose-Einstein condensation(BEC). 
In the first section we introduce very briefly the BEC of harmonically trapped ideal Bose gas. In the second section we theoretically argued that Bose-Einstein(B-E) statistics needs a correction for finite system at ultralow temperatures. As a corrected statistics we introduced a Tsallis type of generalized B-E statistics. The condensate fraction calculated with this generalized B-E statistics, is satisfied well with the experimental result. In the third section we show how to apply the scaling theory in an inhomogeneous system like harmonically trapped Bose condensate at finite temperatures. We calculate the temperature dependence of the critical number of particles by a scaling theory within the Hartree-Fock approximation and find that there is a dramatic increase in the critical number of particles as the condensation point is approached. Our results support the experimental result which was obtained well below the condensation temperature. In the fourth section we concentrate on the thermodynamic Casimir force on the Bose-Einstein condensate. We explored the temperature dependence of the Casimir force. 
\end{abstract}
\maketitle 

\section{Introduction}

                   Bose-Einstein condensation (BEC) is a topic of high experimental \cite{1,2,3,4,5,6,7} and theoretical \cite{8,9,10,11,12} interest. Within the last ten years several thousand works were done on this topic. Below a finite temperature a macroscopic number of Bose-particles come down to the single particle ground state. This phenomenon is called Bose-Einstein condensation. Generally inter particle interaction is responsible for a phase transition. But the BEC type of phase transition occurs entirely due to  Bose-Einstein (BE) statistics. It is a consequence quantum statistical effects of Bose particles. The theory of BEC was given in 1924 by Satyendra Nath Bose and Albert Einstein \cite{12}. But it took 70 years for the experimentalists to verify the theory \cite{1,2,3}. It is a confirmatory test of many body physics. Since 1995 a lot of BEC related experiments are being performed. The experiments opened up a lot of aspects of ultralow temperature physics and threw a lot of challenges to the theoreticians.

                   Since the BEC type of phase transition does not occur due to the inter particle interaction rather it is a consequence of quantum statistical effect, the average separation($\bar{l}$) of the particles is required to be much larger than the s-wave scattering length($a_s$) to probe this quantum statistical effect. Another necessary condition to probe this quantum statistical effect is that $\bar{l}$ is comparable to the thermal de Broglie wave length($\lambda_T$). The $P-T$ diagram of a Bose condensate fall within the solid phase \cite{8}. So, the equilibrium temperatures and pressures in which the Bose condensation is achieved, correspond to the solid phase. Hence the Bose condensate is a different phase other than solid, liquid and gas. It is a metastable state and in equilibrium it becomes a solid. Hence, to achieve Bose condensate one should take a very dilute Bose gas, otherwise the atoms will collide to form a solid. In the very low enough temperature the collision effects are less dominant. So one should go to the very low temperature to achieve the BEC. However, even at the low enough temperature, there is a possibility of two body collision. Due to the two body collision no energy is released out side the two body. Hence there is no possibility to form a molecule due to this type of collisions. In the low enough temperature there is finite possibility of three body collision. Due to this type of collisions the two body can release their energy to a third atom and they can form a molecule and eventually the whole condensate becomes a solid. Even if there is a wall the two atom can release their energy to the wall and the whole system can become a solid. For this reason the atoms are being trapped magnetically . For the trapping, the alkali atoms are good choice for their non vanishing magnetic dipole moment. The experimental studies of BEC started from 1970. The first studies of BEC were focused on the hydrogen atom which was considered because of its light mass, the most natural candidate for realizing BEC. During the experiments hydrogen atoms were first cooled in a dilution refrigerator, then trapped by a magnetic field and farther cooling by evaporation took it close to the BEC. In the 1980s laser-based technique, such as laser cooling and magneto optical trapping were developed to cool and trap the neutral atoms. Alkali atoms are well studied to laser-based methods for their optical transition can be excited by available lasers and for they have a favorable internal energy-level structure to cool to very low temperatures. Once they are trapped their temperatures can be lowered further by evaporative cooling. Following the different cooling techniques the experimental teams of Cornell and Wiemam at Jila and of Ketterle at MIT succeeded in 1995 in reaching the very low temperatures and the densities to observe Bose-Einstein condensation in $^{87}$Rb and $^{23}$Na respectively \cite{1,2}. In the same year the BEC of $^7$Li was also achieved by Hulet group \cite{3}. Successively the BEC of $^1$H \cite{13}, $^4$He \cite{14}, $^{41}$K \cite{15} and $^{52}$Cr \cite{7} was achieved. The BEC of $^{85}$Rb atoms which interacts attractively was also obtained by Jila group in 2001 \cite{16}. In the last few years the experimental and theoretical studies of BEC-BCS crossover draw a lot of attention \cite{17}. Recently the experimental observation of strong quantum depletion in BEC has been achieved by MIT group \cite{18}. Since 1995 the experimental study of BEC is unraveling a lot of new physics in the ultralow temperatures and is throwing a lot of challenges to the theoreticians. 

                  The theory of statistical mechanics as well as the theory of Bose-Einstein statistics and hence the theory of Bose-Einstein condensation is standing on the criteria of thermodynamic limit of the system. In the thermodynamic limit the volume($V$) of the system goes to infinity and the number($N$) of particles of the system goes to infinity such that $\frac{N}{V}$ is finite. However, according to the experimental setup of the BEC the system size is $\sim mm^3$ and the number of particles is $\sim 10^5$ \cite{1,2,3,4,5,6,7}. So, the criteria of thermodynamic limit is not satisfied. Hence we are interested in exploring the finite size corrections to the Bose-Einstein condensation.

\subsection{Theory of Bose-Einstein condensation for trapped ideal Bose gas}

There are two types of particles in the nature: fermions and bosons. Fermions have half integral spins and the bosons have integral spins. The Bose-Einstein condensation(BEC) has been achieved for $^{87}$Rb, $^{23}$Na, $^7$Li, $^4$He, $^2$H, $^{52}$Cr etc. These are not the fundamental bosons like photons, gluons, Z-bosons etc. These alkali atoms are made up of even number of fermions(electrons, protons and neutrons). The spins of these fermions add up to make the atoms composite bosons. BEC is a quantum statistical effect of the bosons. It should be mentioned that the BEC of the photons is not possible because the number of photons is not a conserved quantity. Due to quantum statistics two or more identical bosons prefer to occupy a single quantum state. Below a certain temperature called the condensation temperature($T_o$) a macroscopic number of bosons prefer to occupy the single particle ground state. A condensate is formed with the particles in the single ground state. For this reason the condensate gets a long range order and the condensate behaves like a macroscopic object.

BEC has been experimentally achieved for harmonically trapped Bose gas. For the trapped system the condensation is achieved in the position space as well as in the momentum space. 
Let us consider a 3 dimensional system of harmonically trapped Bose gas, where all the particles are simple harmonic oscillators. Here all the particle are oscillating along a fixed point. Generally the oscillation frequencies along the three perpendicular directions are different. But, for the simplicity of the calculations we consider an isotropic case, where all the frequencies are the same. For the ideal case there are no interparticle interactions between the particles and the statistical mechanical behavior of the system can be well understood by a single particle hamiltonian.

The energy eigenvalues of the single particle hamiltonian is 
\bea
\epsilon_n=(n+\frac{3}{2})\hbar\omega
\eea
where $n=0,1,2,3....$ and $\omega$ is the angular frequency of the oscillators.
 
For the three dimensional harmonic oscillator the degeneracy of the state which has the energy $(n+\frac{3}{2})\hbar\omega$ is $(n^2+3n+2)/2$. The first term of the degeneracy factor is the bulk term and second term is the surface term. In the thermodynamic limit only the bulk term dominates. In the following section we shall show that the surface term will give the finite size correction.
\subsection{Calculation of the condensation temperature($T_o$) and condensation fraction}
We consider that the Bose gas is in equilibrium with its surroundings at temperature $T$.
At and below $T_o$ the chemical potential goes to $(3/2) \hbar\omega$.
The total number of particles below $T_o$ is obtained as
\bea
N_T=\int_0^{\infty}\frac{n^2/2}{e^{n\hbar\omega/kT}-1}dn=[\frac{kT}{\hbar\omega}]^3\zeta(3)
\eea
At $T_o$ there is no particles in the ground state. So, we must have
\bea
N=[\frac{kT_o}{\hbar\omega}]^3\zeta(3)
\eea

From the above equation we get the condensation temperature as
\bea
T_o=\frac{\hbar\omega}{k}[\frac{N}{\zeta(3)}]^{1/3}
\eea
Let us now define the thermodynamic limit of the harmonically trapped Bose gas.
In the thermodynamic limit $\omega\rightarrow 0$ and $N\rightarrow\infty$ such that $N\omega^3$ is a finite constant.
From the eqn.(2) and (3) we get the expression of condensed fraction as
\bea
\frac{N_0}{N}=1-\frac{N_T}{N}=1-[T/T_o]^3
\eea
 
From the above equation we see a macroscopic occupation of the particles to the lowest single particle energy state is possible below a finite temperature $T_o$. This is a consequence of Bose-Einstein statistics and this phenomenon is called Bose-Einstein condensation. 

\subsection{Specific heat}
The total number of particles above $T_o$ is obtained as
\bea
N=\int_0^{\infty}\frac{n^2/2}{z^{-1}e^{n\hbar\omega/kT}-1}dn=[\frac{kT}{\hbar\omega}]^3g_{3}(z)
\eea
From this relation we get
\bea
\frac{dz}{dT}\mid_N=-\frac{3g_3(z)}{Tg_2(z)}.
\eea

The total energy of the gas is obtained as
\bea
E=\int_0^{\infty}n\hbar\omega\frac{n^2/2}{z^{-1}e^{n\hbar\omega/kT}-1}dn=3kT[\frac{kT}{\hbar\omega}]^3g_{4}(z)
\eea
for $T>T_o$.
From the above eqn.(8)and (7) we get the specific heat as 
\bea
C_v=12k[\frac{kT}{\hbar\omega}]^3g_4(z)-9k[\frac{kT}{\hbar\omega}]^3g_3^2(z)/g_2(z)
\eea
At and below $T_o$ we have $z=1$. So the specific heat for $0<T\le T_o$ is
\bea
C_v=12k[\frac{kT}{\hbar\omega}]^3g_4(1)
\eea

From the eqn.(9) and (10) we see that there is a discontinuity in $C_v$ at $T=T_o$. The amount of discontinuity is
\bea
\triangle C_v=-9\frac{\zeta(3)}{\zeta(2)}Nk
\eea
This discontinuity (or the discontinuity in the derivative) in specific heat signalize that BEC is a phenomenon of phase transition. 

\subsection{Shifts of $T_o$}

             In the previous section we stated that the degeneracy factor of a state of the 3-d isotropic harmonic oscillator with energy $(n+3/2)\hbar\omega$ is $(n^2+3n+2)$. In the thermodynamic limit only the first term of the degeneracy contributes. In the following calculations we will see that the second term on the degeneracy factor will give a finite size correction. The corrections due to the third term on the degeneracy factor and that due to summation integral conversion are negligible. 
Now including the second term of the degeneracy factor we get total number of the excited particles below the condensation temperature as  
\begin{eqnarray}
N_T&=&\int_{0}^{\infty}\frac{n^{2}}{2}\frac{1}{e^{n\hbar\omega/kT}-1}\,dn+\int_{0}^{\infty}\frac{3n}{2}\frac{1}{e^{n\hbar\omega/kT}-1}\,dn\nonumber\\&=& (\frac{kT}{\hbar\omega})^{3}\zeta(3)+\frac{3}{2}(\frac{kT}{\hbar\omega})^2\zeta(2)
\end{eqnarray}

Due to the inclusion of the second term of the degeneracy factor we get the new condensation temperature $T_c$ such that

\begin{eqnarray}
N=(\frac{kT_c}{\hbar\omega})^{3}\zeta(3)+\frac{3}{2}(\frac{kT_c}{\hbar\omega})^2\zeta(2)
\end{eqnarray} 

Comparing the eqn.(3) with the eqn.(13) we get a shift of the condensation temperature $T_c-T_o$, such that \cite{19} 
\bea
\frac{\triangle T_o}{T_o}=\frac{T_c-T_o}{T_o}=-\frac{\zeta(2)}{2[\zeta(3)]^{2/3}}N^{-1/3}=-0.73 N^{-1/3}
\eea
From the eqn.(14) we see that the finite size induce a lowering of the condensation temperature.

\section{More accurate theory for Bose-Einstein condensation fraction}
More accurate theory for Bose-Einstein condensation fraction was discussed in \cite{20}. 

In the thermodynamic limit we have the B-E statistics as
\bea
\bar{n_i}=\frac{1}{e^{(\epsilon_i-\mu)/kT}-1}.
\eea
                              
               The mass of each particle be $m$. The length and the volume of the system are $L$ and $V(=L^3)$ respectively. Thermal de-Broglie wave length of a single particle is $\lambda_T=\sqrt{\frac{2\pi\hbar^{2}}{mkT}}$, where $k$ is the Boltzmann constant. Average separation of particles is $\bar{l}=(\frac{V}{N})^{1/3}$. In the classical limit we must have
\bea
\frac{\bar{l}}{\lambda_T}\gg 1.
\eea
From eqn.(16) we can easily write 
\bea
\frac{kT}{\frac{2\pi\hbar^2}{mL^2}}\gg N^{2/3}.
\eea
At sufficiently low temperatures when $\frac{\bar{l}}{\lambda_T}\sim 1$, the gas becomes degenerate and quantum correction is necessary. BEC occur at the onset of this degeneracy. So, for $T\sim T_o$ we have $\frac{kT_o}{\frac{2\pi\hbar^2}{mL^2}}\sim N^{2/3}$. However the condition of thermodynamic limit($L/\lambda_T\rightarrow\infty$) is also valid at this temperature. For finite system the thermodynamic limit is realized as $L/\lambda_T\gg 1$. However, at sufficiently ultralow temperatures $L/\lambda_T$ can be comparable to 1. In this situation, the B-E statistics needs a correction.

             The usual condition of thermodynamic limit is not properly satisfied in the case of the experimental setup of BEC of 3-d harmonically trapped Bose gas\cite{1,2,3,4,5,6,7}. In the semiclassical approximation the number density at the distance(${\bf{r}}$) from the center of the trap is\cite{8} 
\bea
n_T({\bf{r}})=\int_0^{\infty}\frac{1}{e^{(\frac{p^2}{2m}+\frac{m\omega^2r^2}{2})/kT}-1}\frac{4\pi p^2dp}{(2\pi\hbar)^3}=\frac{1}{\lambda_T^3}g_{\frac{3}{2}}(e^{-\frac{m\omega^2r^2}{2kT}}),
\eea
where $\lambda_T=\sqrt{\frac{2\pi\hbar^2}{mkT}}$, and $g_{\frac{3}{2}}(x)=x+x^2/2^{3/2}+x^3/3^{3/2}+...$ is Bose-Einstein function of a real variable $x$. $\bar{n}(r)\sim e^{-\frac{m\omega^2r^2}{2kT}}$, where $\frac{1}{2}m\omega^2r^2$ is the trap potential and $\omega$ is the angular trap frequency. The length scale of this 3-d trapped Bose gas is $L_T\sim \sqrt{\frac{2kT}{m\omega^2}}$. Putting this expression of length in eqn.(17), we get the conditions of classical limit as $\frac{kT}{\hbar\omega}\gg N^{1/3}$, and the thermodynamic limit($L_T/\lambda_T\rightarrow\infty$) as $\frac{kT}{\hbar\omega}\rightarrow\infty$. Eventually we get the condensation temperature for this system as $T_o\sim\frac{\hbar\omega}{k}N^{1/3}$. More precisely the thermodynamic limit for this system is realized as $\omega\rightarrow 0$,$N\rightarrow\infty$ and $N\omega^3=constant$. From the above semiclassical expression of number density we have \cite{8} 
\bea
L_T=\sqrt{\frac{\hbar}{m\omega}}\sqrt{\frac{2\zeta(4)kT}{\zeta(3)\hbar\omega}}.
\eea
With the consideration of thermodynamic limit, total number of excited particles for $T\le T_o$ are $N_T=[\frac{kT}{\hbar\omega}]^{3}\zeta(3)$ and the number of condensed particles are $N_o=N[1-(\frac{T}{T_o})^{3}]$. In the experimental setup the typical value of\cite{7} $N$, $\omega$ and $m$ are of the order of 50000, 2645$s^{-1}$ and 52 amu. With these experimental parameters, at $T_o$ the length scale as expressed in eqn.(19) is $L_T \sim 56\times 10^{-4}mm.$ and $\lambda_{T_o}=2.88\times 10^{-4}mm.$ and their ratio is $L_T/\lambda_T=32.2$. However, for $T\sim\frac{\hbar\omega}{k}$, a large fraction of particle come down to the ground state and the length scale of the system becomes $\sim \sqrt{\frac{\hbar}{m\omega}}\sim 6.76\times 10^{-4}mm$. At these ultralow low temperatures, the thermal de Broglie wavelength becomes comparable to the system size. At these ultralow temperatures the usual theory of statistical mechanics of finite system is not properly valid. So, for $ \frac{\hbar\omega}{k}\lnsim T\lesssim T_o$, we seek an ultralow temperature as well as finite size correction to the B-E statistics.

\subsection{Correction of B-E statistics from qualitative point of view}

               To quantify the correction arising from ultralow temperatures let us start from Tsallis statistics\cite{21}. The relative probability that $E$ be the total energy of a system is given by Boltzmann factor $e^{-E/kT}$. In the Tsallis statistics Boltzmann factor is replaced by $\frac{1}{(1+(q-1)E/kT)^{1/(q-1)}}$, where $q$ is a hidden variable. It is easy to check that as $q\rightarrow 1$, we get back Boltzmann factor. However, for the Bose particles we should start from the Tsallis type of generalized Bose-Einstein statistics which is expressed as\cite{22} 
\bea
\bar{n_i}=\frac{1}{[1+(q-1)\frac{(\epsilon_i-\mu)}{kT}]^{\frac{1}{(q-1)}}-1}.
\eea

Once again it is easy to check that, in the generalized B-E statistics, as $q\rightarrow 1$, we get back B-E statistics. However, Tsallis statistics\cite{21} is applied to equilibrium\cite{23,24,25} as well as to nonequilibrium\cite{26,27,28} systems. Tsallis statistics is re-derived as one of the superstatistics\cite{29} of a nonequilibrium system. In this theory the hidden variable($q$) of Tsallis statistics is equated with system parameter and $q$ is no longer a variable. In the theory of superstatistics\cite{29} $2/(q-1)$ is redefined as effective no. of degrees of freedom. In the theory of dynamical foundation of nonextensive statistical mechanics, this effective no. of degrees of freedom is equated as\cite{28} $(3-q)/(q-1)$. However, for finite equilibrium system we equate $q$ with a system parameter. We will equate $q-1$ with $\lambda_T/L_T$ so that in the thermodynamic limit($L_T/\lambda_T\gg 1$) of finite system we can go to the usual Bose-Einstein statistics. FIG.1 shows experimental and theoretical plot of condensation fraction with temperature. In this figure we see that a negative shift of condensation fraction is necessary to satisfy the theoretical plot with the experimental data. In the theoretical plot there is finite size correction and the correction due to two body interactions. We shall see that the generalized B-E statistics with $(q-1)\propto\frac{\hbar\omega}{kT}$ will give rise to a significant negative shift of condensation fraction with temperature. This significant shift along with finite size correction and the correction due to the two body interaction might satisfy the experimental results. For the trapped Bose gas we shall start from eqn.(20) and put $q-1=\frac{\hbar\omega}{\alpha kT}$ to get 
\bea
\bar{n_i}=\frac{1}{[1+\frac{\hbar\omega}{\alpha kT}\frac{(\epsilon_i-\mu)}{kT}]^{\frac{\alpha kT}{\hbar\omega}}-1},
\eea 
where $\alpha$ is an arbitrary constant. This  $\alpha$ is to be determined from the experimental result. 

\subsection{Corrected and generalized B-E Statistics}
           We are considering a finite system. For this finite system the condensation temperature would not be $T_o$. For this system the condensation temperature is $T_c$ which must be close to $T_o$. For a system of Bose gas in isotropic harmonic trap, the single particle energy levels are $\epsilon_j = (\frac{3}{2}+j)\hbar\omega , \ (j=0,1,2,3,...)$. For $ T\leq T_c$, we have $\mu=\frac{3}{2}\hbar\omega $. So for $T\leq T_c$, from eqn.(21) we get the average number of particles in $j$th state as
\bea
\bar{n_j}=\frac{1}{[1+\frac{j}{\alpha t^{2}}]^{\alpha t}-1}
\eea
where $t=\frac{kT}{\hbar\omega}$. From the above eqn.(22) we get the expression of the total number of excited particles below the condensation point of a 3-D Bose gas isotropic harmonic oscillators as \cite{20}

\begin{eqnarray}
N_T=t^{3}\zeta(3)+\frac{3t^2}{2}\zeta(2)+\frac{6t^2}{\alpha}\zeta(4)
\end{eqnarray}

                The gas being very dilute there should be a correction term in the expression of $N_T$ due to two body scattering. This correction within Hartree-Fock(H-F) approximation has been discussed in \cite{30}. According to H-F approximation, the correction term to the above $N_T$ is $3\times 1.326\frac{a}{\sqrt{\hbar/m\omega}}N_T^{7/6}=4.932\frac{a}{\sqrt{\hbar/m\omega}}t^{7/2}$, where $a$ is the s-wave scattering length. Introducing this term in eqn.(23) we get the more corrected expression of number of excited particle as 
\bea
N_T=t^{3}\zeta(3)+\frac{3t^2}{2}\zeta(2)+4.932\frac{a}{\sqrt{\frac{\hbar}{m\omega}}}t^{7/2}+\frac{6t^2}{\alpha}\zeta(4).
\eea
At $T=T_c$, all the particles will be in the excited states\cite{8,9}. So at $T=T_c$ or at $t_c=\frac{kT_c}{\hbar\omega}$, the number of excited particles is equal to the total number of particles. So, we must have
\bea
N=t_c^{3}\zeta(3)+\frac{3t_c^2}{2}\zeta(2)+4.932\frac{a}{\sqrt{\frac{\hbar}{m\omega}}}t_c^{7/2}+\frac{6t_c^2}{\alpha}\zeta(4)
\eea

\begin{figure}
\includegraphics{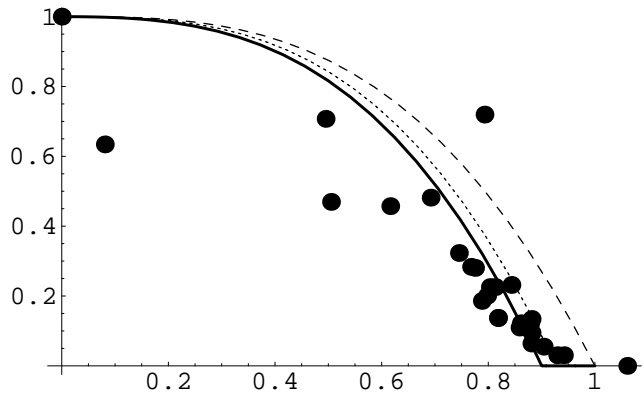}
\caption { Condensation fraction ($\frac{N_o}{N}$) {} versus  temperature($\frac{t}{t_o}$){} plot. The thick line follows from equation(26). The dashed line corresponds to the thermodynamic limit and excludes all the correction terms in equation(26). The dotted line corresponds to the finite size correction and the correction due to interaction excluding the ultralow temperature correction term in equation (26). All the theoretical curves are drawn according to the following experimental parameters. The dotted points are experimental points of Bose-Einstein condensation of $^{52}Cr$ where\cite{7} $N=50000$,$\omega =2645s^{-1}$, $m=52$a.m.u.,$a=105 a_B$ and $T_o\sim 700 nK$} 
\label{fig:Condensation Fraction}
\end{figure}

\subsection{Corrected and more accurate condensation fraction}
In the thermodynamic limit of the ideal trapped Bose gas\cite{8,9} $N_T=t^{3}\zeta(3)$ and its condensation temperature $T_o$ is such that\cite{8,9} $t_o=\left[\frac{N}{\zeta (3)}\right]^{1/3}$, where $t_o=\frac{kT_o}{\hbar\omega}$. Comparing this expression of $t_o$ and $t_c$ of eqn.(25) we see that $t_c<t_o$ and there is a shift $\delta t_c=t_c-t_o$ of condensation temperature due to the inclusion of finite size correction term, correction term due to two body scattering and due to the ultralow temperature correction of B-E statistics. For $t\le t_c$, from eqn.(24) we get the fraction of number of particles in the ground state as
\begin{eqnarray}
\frac{N_o}{N}&=&\frac{N-N_T}{N}=1- [\frac{t}{t_o}]^3-[\frac{3t^2}{2}\zeta(2)+4.932\frac{a}{\sqrt{\frac{\hbar}{m\omega}}}t^{7/2}+\frac{6t^2}{\alpha}\zeta(4)]/[t_o^{3}\zeta(3)]
\end{eqnarray}
\subsection{More accurate $T_o$ shift}
However, in the thermodynamic limit with no inter particle interaction, the expression of this fraction would be\cite{8,9} $[\frac{N_o}{N}]_{T-L}=1-[\frac{t}{t_o}]^3$. Now, from eqn.(26), due to the correction terms we get the fractional change in condensation temperature as 
\begin{eqnarray}
\frac{\delta T_c}{T_o}=-\frac{\zeta(2)}{2[\zeta(3)]^{2/3}}N^{-1/3}-1.326\frac{a}{\sqrt{\hbar/m\omega}}N^{1/6}-\frac{2\zeta(4)}{\alpha[\zeta(3)]^{2/3}}N^{-1/3}
\end{eqnarray}

                From the first term of the eqn.(27), we get the $T_c$ shift due to the finite size correction as \cite{19} $\frac{\delta t_{c}^{f-s}}{t_o}=-\frac{\zeta(2)}{2[\zeta(3)]^{2/3}}N^{-1/3}=-.728N^{-1/3}$ and for 50000 particles we get $\frac{\delta t_{c}^{f-s}}{t_o}=-1.97\%$. From the second term of the eqn.(27), we get the $T_c$ shift due to the correction of two body interaction as \cite{8,9,30} $\frac{\delta t_{c}^{int.}}{t_o}=-1.326\frac{a}{\sqrt{\hbar/m\omega}}N^{1/6}=-6.61\%$ for $\omega=2645s^{-1}$, $a=105 a_B$ and for N=50000 \cite{7}. For this setup $t_o=[\frac{N}{\zeta(3)}]^{1/3}=34.65$. From the third term of the eqn.(27) we get the $T_c$ shift due to ultralow temperature correction of B-E statistics as $\frac{\delta t_{c}^{ult}}{t_o}=-\frac{2\zeta(4)}{\alpha[\zeta(3)]^{2/3}}N^{-1/3}=-\frac{1}{\alpha}5.19\%$ for 50000 particles. According to the experiment $\frac{\delta t_c}{t_o}$ should be $10\%$. This $10\%$ $t_c$ shift is achieved if we put $\alpha=1.48$ in eqn.(27).
 
\begin{figure}
\includegraphics{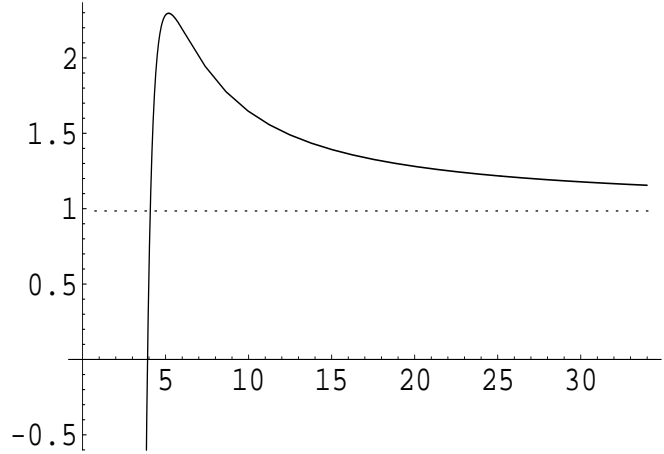}
\caption {Plot of $r'(t)$ with Temperature($t$) in units of $\frac{KT}{\hbar\omega}${}. This plot is followed from eqn.(29). The doted line shows the theoretical value of the ratio $r'(t)$ in thermodynamic limit when $\omega\rightarrow 0$.} 
\label{fig:RATIO OF SPECIFIC HEATS}
\end{figure}
\vskip0.2cm
\subsection{Comparison between the specific heats calculated by the generalized B-E statistics and B-E statistics for the Trapped Ideal Bose Gas}
               Well below $T_c$, as $t$ approaches to 1, the B-E statistics needs more corrections. Below $T_c$, as a comparative study of the two statistics, let us calculate the ratio of specific heat from the two statistics. As a comparative study we can take ideal gas. The total energy of the system would be 
\begin{eqnarray}
E(t) &\sim& (\hbar\omega)\int_{0}^{\infty}\frac{j^{3}}{2}\frac{1}{[1+
\frac{j}{\alpha t^{2}}]^{\alpha t}-1}\,dj \nonumber\\
&=& 3(\hbar\omega)\sum_{i=1}^{\infty}(\alpha t^{2})^{4}\frac{\Gamma (\alpha it -4)}{\Gamma (\alpha it)}
\end{eqnarray}
As $t\gg 1$, $E(t)\rightarrow 3\hbar\omega\zeta (4)t^{4}$ as expected from the B-E statistics\cite{8,9}.

Let us denote the ratio of specific heat($C_v(t)=\frac{k}{\hbar\omega}\frac{d}{dt}E(t)$) calculated from the corrected B-E statistics and from the B-E statistics as 
\bea
r'(t)=\frac{C_v}{k12\zeta(4)t^{3}} 
\eea
Obviously as $t\rightarrow \infty$, $r'\rightarrow 1$. With $\alpha=1.48$ numerical plot of $r'(t)$  at very low temperatures ($1\lnsim t\le t_c$) is shown in the FIG. 2.

         In  FIG. 2 we see that at high temperature $t\gg 1$, the specific heat behaves well according to our familiar $T^{3}$ law. In this figure we also see that specific heat for this finite system becomes negative at $t\lesssim 5$. In this range of temperatures $L_T\lesssim\lambda_T$ and the theory of statistical mechanics is not valid. In spite of that, due to nonextensivity the appearance of negative specific at these temperatures is not surprising theoretically\cite{31} and experimentally\cite{32}. 
\subsection{Conclusion}
           That B-E statistics needs a correction for finite system at ultralow temperatures is justified theoretically. But we did not theoretically justify why Tsallis type of generalized B-E is necessary for finite system at ultralow temperatures. However, Tsallis type of generalized B-E statistics with our redefined parameter satisfy experimental result. We argued that the correction over the B-E statistics is $\it{O}(\lambda_T/L_T)$. Smaller the ratio smaller is the correction. However, this ultralow temperature correction is valid only for a finite system. How $\alpha$ is to be determined theoretically remains an open question. 

\section{Finite temperature scaling theory for Bose-Einstein condensate}
                 Finite temperature scaling theory for Bose-Einstein condensate was well discussed in \cite{33}. 
                 For the ultracold gas the interaction is characterized by s-wave scattering length $a_s$. Atomic interaction as well as the value of scattering length($a_s$) can be controlled in Feshbach resonance\cite{34}. Stability and collapse of Bose gas with negative scattering length has been observed in the clouds of ultracold $^7$Li\cite{35} and $^{85}$Rb\cite{36,37}. If the interaction is attractive ($a_s<0$), the gas tends to increase the density of the central region of the trap. This tendency is opposed by the zero-point energy and thermal energy of the atoms. If the number of atoms is greater than a critical number($N_c$), the central density increases strongly and the zero-point and thermal energy are no longer able to avoid the collapse of the gas. 

                 As the temperature($T$) goes to zero, all the particles come down to the ground state and the system is well described by the ground state wave function $\Psi_0({\bf{r}})=\sqrt{\frac{N}{l^3\pi^{3/2}}}e^{-\frac{r^2}{2l^2}}$ in the position({\bf r}) space, where $l=\sqrt{\hbar/m\omega}$ is the length scale of the oscillators. At $T=0$, the density of the condensed particles is described as $n_0({\bf{r}})=\mid\Psi_0({\bf{r}})\mid^2=\frac{N}{l^3\pi^{3/2}}e^{-\frac{r^2}{l^2}}$. In absence of collision the number density of the excited particles is\cite{8} $n_T({\bf{r}})=\frac{1}{\lambda_T^3}g_{\frac{3}{2}}(e^{-\frac{m\omega^2r^2}{2k_BT}})$, where $\lambda_T=\sqrt{\frac{2\pi\hbar^2}{mk_BT}}$, and $g_{\frac{3}{2}}(x)=x+x^2/2^{3/2}+x^3/3^{3/2}+...$ is the Bose-Einstein function of a real variable $x$.

               Now, if we allow two particle attractive interaction, the particles come closer and the length scale of the system would reduce by a factor $\nu$ ($<1$) such that the ground state wave function would be $\Psi_0(r)={\sqrt{\frac{n_0}{\nu^2 l^3\pi^{3/2}}}}e^{-r^2 /2 \nu^2 l^2}$ and the thermal density of particles would be $n_T({\bf{r}})=\frac{1}{\nu^3 \lambda_T^3}g_{\frac{3}{2}}(e^{-\frac{m\omega^2r^2}{2\nu^2 k_BT}})$.

Let us consider the interaction potential as $V_{int}({\bf{r}})=g\delta^3({\bf{r}})$, where $g=-\frac{4\pi\hbar^2a}{m}$ is the coupling constant and $a=-a_s$ is the absolute value of the s-wave scattering length. For a dilute gas we must have $\frac{a}{l}\ll 1$. The typical two body interaction energy for $N$ number of particles is $\sim N^2 g/2l^3$. For this interaction, the gas tends to increase the density of the central region of the trap. Well below the BEC temperature($T_c$), i.e. for $T\rightarrow 0$, this tendency is resisted by the zero-point energy ($\sim N\hbar\omega$) of the atoms. In the critical situation the typical oscillator energy must be comparable to the typical interaction energy. So, at $T=0$, we must have $N_c\hbar\omega\sim N_c^2 g/2l^3$. From this relation we can write $\frac{N_ca}{l}\sim 1$.
      
                  For $0<T<T_c$, the typical total energy of the system is $N^{4/3}\hbar\omega$. So at these temperatures we  must have $N_c^{4/3}\hbar\omega\sim N_c^2 g/2l^3$. From this relation we have $\frac{N_ca}{l}\sim[\frac{l}{a}]^{1/2}>1$.

                 However, near $T_c$ the length scale of the system is\cite{8} $L_{T_c}\sim l\sqrt{\frac{k_BT_c}{\hbar\omega}}\sim lN^{1/6}$. So, near the condensation temperature we must have $N_c^{4/3}\hbar\omega\sim N_c^2 g/2L_{T_c}^3$. From this relation we have $\frac{N_ca}{l}\sim [\frac{l}{a}]^5\gg 1$.

                 In the following subsection we shall explicitly calculate $\frac{N_ca}{l}$ by a scaling theory  within the Hartree-Fock(H-F) approximation. We shall explicitly show the temperature dependence of the critical number($N_c$). 

\subsection{Energy expression of the Bose particles within Hartree-Fock approximation} 
              Although the problem of an attracting Bose gas was discussed by many authors\cite{9,10}, yet the temperature dependence of the critical number of particles  has not been explicitly explored. H-F approximation for Bose gas has been discussed in \cite{8}.  For this interacting Bose gas, the average occupation number $\bar{n}_i$ as well as the single particle wave functions $\phi_i$ are determined\cite{8} by minimizing the grand potential. Since the entropy and the total number of particles depend on the occupation number $n_i$, the single particle wave functions $\{\phi_i\}$ are simply obtained by minimizing the energy functional with proper normalization constraint $\int d^3{\bf{r}}\mid\phi_i\mid^2=1$ for each $i$\cite{8}. Within the H-F approximation we have the expression of energy functional as\cite{8} 
\begin{eqnarray}
E&=&\int d^3{\bf{r}}[\frac{\hbar^2}{2m}n_0\mid \nabla\phi_0\mid^2+\sum_{i\neq 0}\frac{\hbar^2}{2m}n_i\mid \nabla\phi_i\mid^2\nonumber\\&&+V(r)n_0({\bf{r}})+V({\bf{r}})n_T({\bf{r}})+\frac{g}{2}n_0^2({\bf{r}})\nonumber\\&&+2gn_0({\bf{r}})n_T({\bf{r}})+gn_T^2({\bf{r}})]
\end{eqnarray}
\subsection{scaling argument}
In our scaling theory we introduce a variational parameter $\nu$ which would fix the width of the system as well as would take a proper choice of $n_0({\bf{r}})$ and $n_T({\bf{r}})$. So, in this method, the choice of $\{\phi_i\}$ is denoted by the variational parameter $\nu$. The minimum of the energy functional for a certain choice of $\{\phi_i\}$ would corresponds to the equilibrium of the system. So, for equilibrium condition of the system, the energy functional would be minimized with respect to the variational parameter($\nu$). In FIG. 3 we shall see that above a critical number of particles($N_c$), the energy functional will have no stable minimum. If the energy functional as well as the grand potential has no stable minimum, the grand potential as well as the energy functional would arbitrarily decrease until the width($\nu$) of the system becomes zero. Under this condition the system is said to be collapsed. 

           The scaled form of $n_T({\bf{r}})$ is obtained from the variation of B-E statistics such that $\bar{n}({\bf{p}},{\bf{r}})=\frac{1}{e^{(\frac{p^2\nu^2}{2m}+\frac{m\omega^2r^2}{2\nu^2})/k_BT}-1}$, where {\bf p} is the momentum of a single particle. With this variational form of statistics we have the total number of excited particles as $N=(\frac{k_BT_c}{\hbar\omega})^3\zeta(3)$. For $T\le T_c$, the total number of particles in the ground state is $n_0=N-N_T=N(1-\frac{T}{T_c})^3$. From the variation in statistics we evaluate the energy functional of eqn.(30) in units of $N\hbar\omega$ as\cite{33}
\begin{eqnarray}
X_t(\nu)=c_1(t)(\frac{1}{\nu^2}+\nu^2)-\frac{c_2(t)}{\nu^3}
\end{eqnarray}
where $t=T/T_c$, $c_1(t)=\frac{3}{4}(1-t^3)+\frac{3}{2}\frac{N^{1/3}\zeta(4)t^4}{[\zeta(3)]^{4/3}}$ and $c_2(t)=\frac{1}{\sqrt{2\pi}}\frac{Na}{l}[1-t^3]^2+\sqrt{\frac{8\zeta(3/2)}{\pi}}\frac{N^{1/2}a}{l}t^{3/2}[1-t^3]+S'\sqrt{\frac{2}{\pi[\zeta(3)]^3}}\frac{N^{1/2}a}{l}t^{9/2}$ such that $S'=\sum_{i,j=1}^{\infty}\frac{1}{(ij)^{3/2}(i+j)^{3/2}}\approx 0.6534$. Putting $\frac{a}{l}=0.0066$ \cite{16} and $t=0.8$, in eqn.(31) we get the FIG. 3 for various number of particles. In FIG. 3 we see that $X_t(\nu)$ has a stable minimum and an unstable maximum below a critical number of particles. At the critical number of particles the minimum and maximum coincides. Above the critical number of particles there is no stable minimum and its energy arbitrarily decreases to $-\infty$ and its width becomes zero which signalize the collapse of the gas.

\begin{figure}
\includegraphics{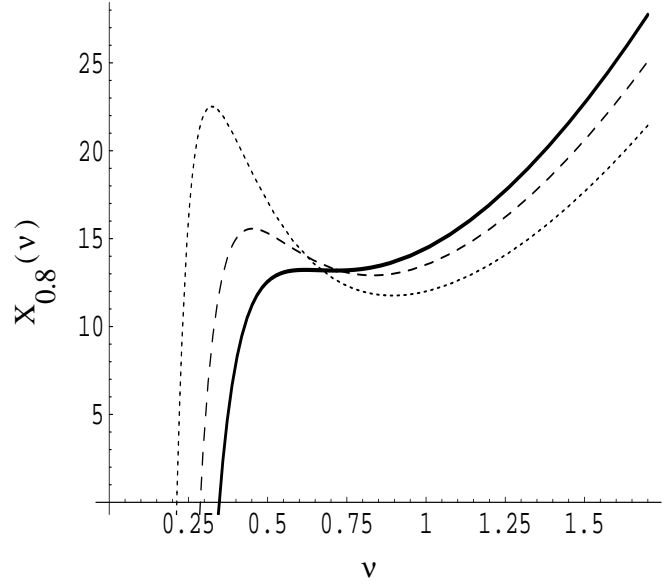}
\caption {Plot of energy per particle ($X_{0.8}(\nu)$)in units $\hbar\omega$ with the variational parameter $\nu$. For the thick line, $\frac{Na}{l}=27.78$. For the dashed line, $\frac{Na}{l}=20$. For the dotted line, $\frac{Na}{l}=12$. All the lines follow from eqn.(31).}
\label{fig:Plot of Energy functional}
\end{figure}
\subsection{Evaluation of the critical number}

            For the critical number of particles we must have $\frac{\partial X_t}{\partial\nu}=\frac{\partial^2 X_t}{\partial\nu^2}\mid_{\nu_c}=0$. From these critical conditions we get the expression of $\frac{N_ca}{l}$ for $t=0$ as 

\begin{eqnarray} 
\frac{N_ca}{l}=\frac{\sqrt{2\pi}}{2}[\nu_c-\nu_c^5]=0.671
\end{eqnarray}
The same result was also obtained in \cite{39}. However, the experimental value is 0.459 \cite{16}.
From the same critical conditions we get the expression of $\frac{N_ca}{l}$  for $0<T<T_c$ as 

\begin{eqnarray}
\frac{N_ca}{l}&=&1.210[\frac{l}{a}]^{1/2}\frac{t^6}{(1-t^3)^3}+\frac{1.006}{[1-t^3]}-[10.666 t^{3/2}\nonumber\\&&\times(1-t^3)+1.636t^{9/2}][\frac{a}{l}]^{1/4}\frac{t^3}{[1-t^3]^{7/2}}
\end{eqnarray}
Eqn.(33) as well as FIG. 4 represent the relation between the critical number of particles and the temperature. However, from this figure and according to eqn.(33) we get $\frac{N_ca}{l}=27.89$ for $t=0.8$. From the FIG. 3 we see that $\frac{N_ca}{l}=27.78$ for $t=0.8$. 
\begin{figure}
\includegraphics{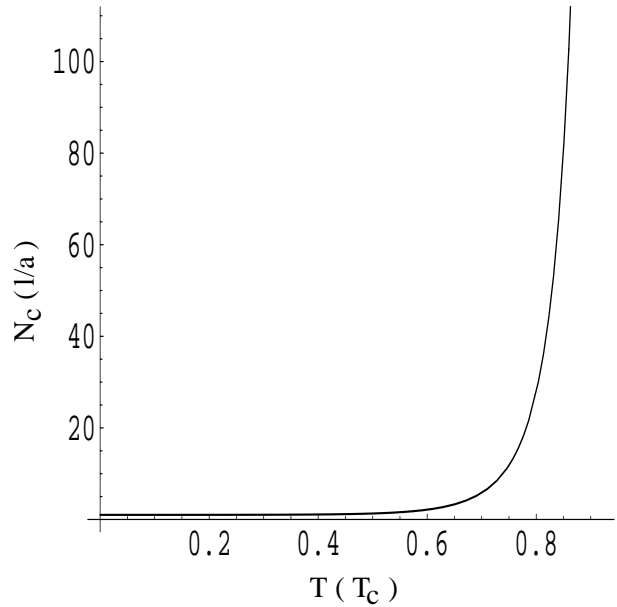}
\caption {Plot of the critical number $N_c$ in units of $\frac{l}{a}$ with temperature($T$) in units of ($T_c$). Here $\frac{l}{a}= 0.0066$ \cite{16}. This plot follows form eqn.(33). }
\label{fig:Critical Number of Particles with Temperature}
\end{figure}

            The same critical conditions give the expression of $\frac{N_ca}{l}$  for $T=T_c$ as

\begin{eqnarray}
\frac{N_ca}{l}=2.253[\frac{l}{a}]^5.
\end{eqnarray}
Comparing the eqn.(33) and eqn.(34), we see that there is a dramatic increase of $N_c$ as the condensation point is approached. It would be interesting for the experimentalists to do the experiment near $T_c$ and to verify this dramatic increase of the exponents of $a/l$ in the expression of $N_c$.

\subsection{Conclusion}
                  In the FIG. 4, we see that the temperature dependence of $N_c$  is not significant for $0\le T\lesssim 0.5 T_c$. If the experiments of the collapse is performed within this range of temperature, the zero temperature theory should well satisfy the experimental results. Here we have assumed that the radius of the thermal cloud changes by the same factor as that of the condensate cloud due to the fact that $L_T$ is proportional to $l$. Initially we qualitatively  explained the physics of the collapse of the attractive atomic Bose gas. Then we semi qualitatively estimated $N_ca/l$ for the various range of temperatures ($0\le T\le T_c$). Finally we calculated $N_ca/l$ by a scaling theory (variational method) within the H-F approximation. Our calculation supports the  semi qualitative estimation of $N_ca/l$. It proves that exponent of $a/l$ in the expressions of $N_c$ for the different range of temperature does not depend on the scaling assumption. The same temperature dependence of $N_c$ as well as the same type of exponents can be obtained by a quantum kinetic theory. 
   
\section{Casimir effect on Bose-Einstein condensate confined between two slabs}         
Casimir effects on Bose-Einstein condensate confined between two slabs were well discussed in \cite{40}. Vacuum fluctuation of electromagnetic field would cause an attractive force between two closely spaced parallel conducting plates. This phenomenon is called Casimir effect and this force is called Casimir force \cite{41,42,43}. Before going into details we first estimate the Casimir force intuitively. Since this force causes due the vacuum(quantum) fluctuations of the electromagnetic field there will be a term $\hbar$ in the expression of vacuum energy. As we are considering the vacuum of electromagnetic field, there must be a term {\bf c}(velocity of light) in the expression of vacuum energy. Since the quantized energy of the electromagnetic field is $\hbar$ times frequency we can conveniently write the expression of vacuum energy as $\hbar c/L$, where $L$ is the plate separation. Since the fluctuation are confined between the plate, the Casimir energy is expected to be proportional to the area($A$) of the plate. For the proper dimension, the Casimir energy would be $\sim A \hbar c/L^3$. So the Casimir energy would be $\sim A \hbar c/L^4$. This is an estimation of the Casimir force. In the following we show an exact calculation of the Casimir force.

             Let the plates be kept parallel to the x-y plane. They are separated $L$ along the z- axis. The vacuum energy of an electromagnetic field is $E=\Sigma_{\bf k}\frac{1}{2}\hbar\omega_{\bf k}$, where $\omega_{\bf k}$s are the angular frequencies of the photons. The vacuum energy, in term of the wave number($k_x,k_y,k_z$) can be written as 
\bea
E(L)= 2A\times\frac{1}{2(2\pi)^2}\hbar c \Sigma_{n=0}^{\infty}\int_{0}^{\infty}\sqrt{k_{\perp}^2+\frac{n^2\pi^2}{L^2}}2\pi k_{\perp}dk_{\perp}
\eea 
where $k_{\perp}^2=k_x^2+k_y^2$ and $k_z= n\pi/L$. The factor 2 of the above equation comes due to the fact that photon has two polarizations. With the substitution $k_{\perp}^2+\frac{n^2\pi^2}{L^2}=t^2$ and $\Lambda$ as the cutoff of $t$ we can recast the eqn.(35) as 
\bea
E(L)=A\frac{2\pi\hbar c}{(2\pi)^2}\Sigma_{n=0}^{\infty}\int_{\frac{n\pi}{L}}^{\Lambda}t^2dt=\frac{\hbar c A}{2\pi}[\frac{\Lambda^3}{3}- \frac{n^3\pi^3}{3L^3}].
\eea

From the definition of Casimir force ($F_c(L)=-\frac{\partial}{\partial L}[E(L)-E(\infty)]$)
we get the expression of Casimir force as 
\bea
F_c(L)= -\frac{A\hbar c \pi^2}{2L^4}\Sigma_{n=0}^{\infty}n^3=-\frac{A\hbar c \pi^2}{2L^4}\zeta(-3)=-\frac{A\pi^2\hbar c}{240 L^4}
\eea
where $\zeta(-3)=\frac{1}{120}$ is obtained from the analytic continuation technique. This expression of the Casimir force is valid only at zero temperature. This force has been measured experimentally \cite{44,45}.

         However, Casimir effect can be generalized \cite{46} for any range of temperature and for any dielectric substance  between two dielectric plates. It has also been generalized for thermodynamical systems \cite{43}. Casimir force for this kind of systems has recently been measured \cite{47}. At finite temperature $T$, the definition of Casimir force is generalized as \cite{48}
\bea
F_c(T,L)=-\frac{\partial}{\partial L}[\Omega_T(L)-\Omega_T(\infty)]
\eea
where $\Omega_T(L)$ is the grand potential of the system confined between two plates separated at a distance $L$.

               We consider the Casimir effect for a thermodynamical system of Bose gas between two infinite slabs. Geometry of the system on which some external boundary condition can be imposed is responsible for the Casimir effect. Thermalized photons (massless bosons) in between two conducting plates of area $A$ at temperature $T$ gives rise to the Casimir pressure  \cite{49}
\begin{eqnarray}
\frac{F_c(L)}{A}\sim &-&\frac{\pi^{2}\hbar c}{240L^{4}}[1+\frac{16(kT)^{4}L^{4}}{3(\hbar c)^{4}}]\   \ for\  \frac{\pi\hbar c}{kTL}\gg 1 \nonumber\\ \sim &-&\frac{kT\zeta(3)}{8\pi L^{3}}\    \ for\  \frac{\pi\hbar c}{kTL}\rightarrow 0
\end{eqnarray}
where $k$ is the Boltzmann constant, $c$ is the velocity of light and $L$ is the separation of the parallel plates. At $T\rightarrow 0$, Casimir pressure becomes $-\frac{\pi^{2}\hbar c}{240L^{4}}$ and it is only the vacuum fluctuation which contributes to the Casimir pressure. At high temperature i.e. for $\frac{\pi\hbar c}{kTL}\rightarrow 0$, the Casimir force for photon gas  goes as $L^{-3}$ and has a purely classical expression independent of $\hbar$. 

\subsection{Calculation of the B-E condensation temperature}
              Let us consider a Bose-gas confined  between two infinitely large square shaped hard plates of area $A$. The plates are along x-y plane and they are separated along z- axis by a distance $L$. For the slab geometry, $\sqrt{A}\gg L$. We consider that our system is in thermodynamic equilibrium with its surroundings at temperature $T$. At this temperature the thermal de Broglie wavelength of a single particle of mass $m$ is $\lambda=\sqrt{\frac{\pi\hbar^{2}}{2mkT}}$. In the thermodynamic limit, $\frac{\lambda}{L}\ll 1$. For this system the single particle energy is $\epsilon(p_x,p_y,j)=\frac{p_x^2}{2m}+\frac{p_y^2}{2m}+\frac{\pi^2\hbar^2 j^2}{2mL^2}$, where $p_x$ and $p_y$ are the momentum along x-axis and y-axix respectively and $j=1,2,3,....$. However in the thermodynamic limit the single particle energy can be written as $\epsilon(p_x,p_y,p_z)=\frac{p_x^2}{2m}+\frac{p_y^2}{2m}+\frac{p_z^2}{2m}$, where $p_z$ is the momentum along z-axis.

             Considering the thermodynamic limit the total number of thermally excited particles can be written as
\bea
N_T=\int_{-\infty}^{\infty}\int_{-\infty}^{\infty}\int_{-\infty}^{\infty}\frac{1}{e^{\frac{[\frac{p_x^{2}}{2m}+\frac{p_x^{2}}{2m}+\frac{p_z^{2}}{2m}-\mu]}{kT}}-1}\frac{Vdp_xdp_ydp_z}{[2\pi\hbar]^{3}}
\eea
where $\mu$ is the chemical potential and $V$ is the volume of the system. For this system the BEC temperature can be calculated as
\bea
T_o=\frac{1}{k}[\frac{2\pi\hbar^{2}}{m}][\frac{N}{V\zeta(3/2)}]^{\frac{2}{3}}
\eea
       
\subsection{Finite size correction to the grand potential}
                Let us now introduce the finite size correction. The ground state energy of our system is $[g=\frac{\pi^{2} \hbar^{2}}{2mL^{2}}]$. The average no. of particles with energy $\epsilon_{p_x,p_y,j}$ is given by $\frac{1}{e^{[\frac{p_x^2}{2m} +\frac{p_y^2}{2m}+\frac{\pi^{2} \hbar^{2}(j^{2}-1)}{2mL^{2}}-\mu']/kT}-1}$ where $\mu'=(\mu-g)\le 0$ for bosons. At and below the condensate temperature $\mu'\rightarrow 0$. For this bosonic system we have the grand potential 

\begin{eqnarray}
&&\Omega(A,L,T,\mu')\nonumber\\&&=-kT\sum_{i=1}^{\infty}\int_{p_x=0}^{\infty}\int_{p_y=0}^{\infty}\sum_{j'=0}^{\infty}\frac{Adp_xdp_y}{[2\pi\hbar]^2}\nonumber\\&&\frac{e^{\frac{i\mu'}{kT}}e^{-\frac{ip_x^2}{2mkT}}e^{-\frac{ip_y^2}{2mkT}}e^{-i(\pi(\frac{\lambda}{L})^{2}[j'^{2}+2j'])}}{i}
\end{eqnarray}
where $\lambda =\sqrt{\frac{\pi\hbar^{2}}{2mkT}}$.
Integrating over $p_x$ and $p_y$ we get
\begin{eqnarray}
\Omega(A,L,T,\mu')=-\frac{A(kT)^2 m}{2\pi\hbar^2}\sum_{i=1}^{\infty}\sum_{j'=0}^{\infty}\frac{e^{i\mu'/kT}}{i^{2}}[e^{-\frac{\pi i \lambda^{2}j'^{2}}{L^{2}}}][1-2j'\frac{\pi i \lambda^{2}}{L^{2}}+2j'^{2}(\frac{\pi i \lambda^{2}}{L^{2}})^{2}-\frac{4}{3}j'^{3}(\frac{\pi i \lambda^{2}}{L^{2}})^{3}+...]
\end{eqnarray}
Since $\frac{\lambda}{L}\ll 1$, higher order terms of the above series would not contribute significantly. From Euler-Maclaurin summation formula we convert the summation over $j'$ to integration. The Euler-Maclaurin summation formula for a smooth function $f(k)$ is denoted as 
\bea
\Sigma_{k=0}^{\infty}f(k)=\int_{k=0}^{\infty}f(k)+\frac{1}{2}f(0)-\frac{1}{12}f'(0)+\frac{1}{720}f'''(0)-\frac{1}{30240}f'''''(0)+....
\eea
Now from eqn.(43) and eqn.(44) we have \cite{40}

\begin{eqnarray}
&&\Omega(A,L,T,\mu')\nonumber\\&&=-\frac{A(kT)^2m}{2\pi\hbar^2}\sum_{i=1}^{\infty}\frac{e^{i\mu'/kT}}{i^{2}}[\frac{L}{2\lambda i^{1/2}}-\frac{1}{2}+\nonumber\\&&\frac{\pi}{2}i^{1/2}\frac{\lambda}{L}+\it{O}([\frac{\lambda}{L}]^{2})]\nonumber\\&&=-\frac{A(kT)^2m}{2\pi\hbar^2}[\frac{L}{2\lambda}g_{\frac{5}{2}}(z)-\frac{1}{2}g_{2}(z)+\frac{\pi\lambda}{2L}g_{\frac{3}{2}}(z)]\nonumber\\
\end{eqnarray}
where $z=e^{\mu'/kT}$ is the fugasity and $g_l(z)=z+\frac{z^2}{2^{l}} +\frac{z^3}{3^{l}}+....$ is the Bose-Einstein function. From the above equation we get the total number of particles as 

\begin{eqnarray}   
&&N=-\frac{\partial\Omega}{\partial\mu'}\nonumber\\&&=\frac{AkTm}{2\pi\hbar^2}[\frac{L}{2\lambda}g_{\frac{3}{2}}(z)-\frac{1}{2}g_{1}(z)+\frac{\pi\lambda}{2L}g_{\frac{1}{2}}(z)]
\end{eqnarray}
In the thermodynamic limit of a system, as $T\le T_o$, $z\rightarrow 1$. For a finite system this can not happen, otherwise the correction terms in the above expression would be infinite. Instead at $T\gtrsim T_o$, $z\sim 1$. Taking only the first correction term in the eqn.(46) we have $g_1(z)=-ln(1-z)=[N'(T)g_{\frac{3}{2}}(z)-N\zeta(3/2)]\frac{L}{\lambda N'(T)}=-ln\triangle z$, where $N'(T)= \frac{AkTm}{2\pi\hbar^2}[\frac{L}{2\lambda}\zeta(3/2)]$ and $\triangle z=1-z$ is a small change in the fugasity at $T\gtrsim T_o$. Now putting $z=1$ in the expression of $\triangle z$, we get $\triangle z=e^{-\triangle N\zeta(3/2)L/(N'(T)\lambda)}$, where $\triangle N=N'(T)-N$. We see that in the thermodynamic limit($L\rightarrow\infty$) $\triangle z=0$ and when $L$ is finite such that $L/\lambda\gg 1$, we have $z\sim 1$ at $T\gtrsim T_o$. 

\subsection{Calculation of the Casimir force}

               Let us now calculate the Casimir force. At $T\le T_o$ we put $\mu'\rightarrow 0$ or $z\rightarrow 1$. Now, from the eqn.(45),(41) and (38) we get the expression of Casimir force for $0\le T \le T_c$ as \cite{40}

\bea
F_c(T,L)=-N[\frac{T}{T_o}]^{3/2}\frac{\pi^{2}\hbar^{2}}{mL^{3}}
\eea

            Above the condensation temperature $\mu'<0$ or $z<1$. However, for $T\gtrsim T_o$, $z\sim 1$. Putting $T=T_o+\triangle T$ in equation (45) and from the definition of Casimir force we have \cite{40}
\bea
F_c(T,L)\approx-N\frac{\pi^{2}\hbar^{2}}{mL^{3}}
\eea 
where $0<\frac{\triangle T}{T_o}\ll 1$. Hence, for $T\gtrsim T_o$, the Casimir force weakly depends on temperature.

\begin{figure}
\includegraphics{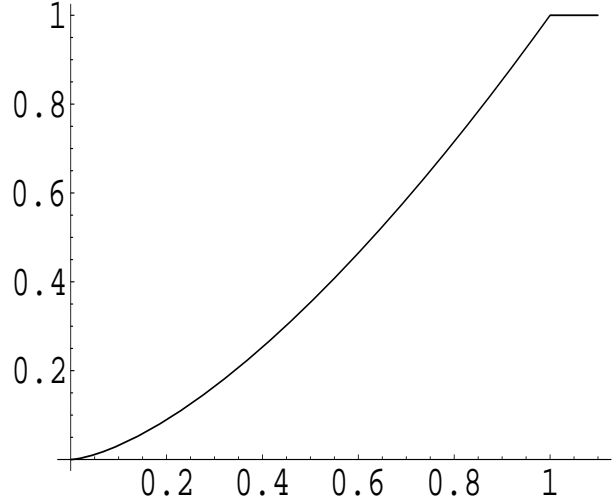}
\caption {Casimir force($F_c(T,L)$) versus temperature($T$) plot. $F_c(T,L)$ is in units of $N\frac{\pi^2\hbar^2}{mL^3}$ and $T$ is in units of $T_o$. Equation (47) corresponds to the range $\frac{T}{T_o}\le 1$. Equation (48) corresponds to the range $1<\frac{T}{T_o}\le 1.1$.} 
\label{fig:Casimir Force}
\end{figure} 
\subsection{Casimir Force for $T\gg T_o$}
            Let us now calculate the Casimir force at $T\gg T_o$. At these temperatures $z\ll 1$. So we can approximately write $g_i(z)\approx z+\frac{z^2}{2^i}$. From the first term of eqn.(46) we have $g_{\frac{3}{2}}(z)=\frac{N2\pi\hbar^2 2\lambda}{AkTmL}\approx z+z^2/(2\sqrt{2})$. For this range of temperatures we can write $e^{\mu/kT}=ze^{-\pi\lambda^2/L^2}\approx z$. For convenience, we replace $\mu'$ by $\mu$ and recast equation (43) as \cite{40}
\begin{eqnarray}
\Omega(A,L,T,\mu)=-\frac{A(kT)^2m}{2\pi\hbar^2}\sum_{i=1}\sum_{j=1}\frac{e^{\mu i/kT}}{i^2}(\frac{L}{2\lambda i^{\frac{1}{2}}}-\frac{1}{2}+\frac{L}{\lambda i^{\frac{1}{2}}}e^\frac{-\pi L^2j^2}{i\lambda^2})
\end{eqnarray}
where we use the formula $\sum_{n=-\infty}^{\infty}e^{-\pi an^2}=\frac{1}{\sqrt{a}}\sum_{n=-\infty}^{\infty}e^{-\pi n^2/a}$. For $\frac{T}{T_o}\rightarrow\infty$, in the expression of above Casimir potential(the third term of the grand potential) we can  put $e^{\mu/kT}=z\ll 1$ and can take $i=1$ and $j=1$ as the leading term to contribute in the Casimir potential. So, for $T\gg T_o$ we have the Casimir force as \cite{40}
\begin{eqnarray}
F_c(T,L)=-\frac{8N(kT)^2mL}{\hbar^2}e^{-\frac{2mL^2kT}{\hbar^2}}.
\end{eqnarray}
   
Now we see that in the classical limit($T\gg T_o$) the Casimir force vanishes as $e^{-kT}$. It is interesting to see that the Casimir force for $T\gg T_o$ is long ranged(power law decay) and for $T\le T_o$ is short ranged(exponential law decay).

             The changes of Casimir force with temperature for the range $0<T\le T_o$ and for the range $T\gtrsim T_o$ is shown in FIG. 5. 
\subsection{Casimir force for trapped geometry}
Casimir effect on BEC for trapped geometry was discussed in \cite{50}. We consider a system of ideal Bose gas which behaves as isotropic trapped harmonic oscillators along two perpendicular directions and along the other perpendicular direction it behaves like particles is 1-d box. At thermodynamic limit the system size must be greater greater than the thermal de Broglie wavelength $\sqrt{\frac{\pi\hbar^{2}}{2mkT}}$ of the particles. For a slab geometry of a 3-d system the length scale along one direction is much less than that along the other two perpendicular directions. Similar type of analysis for this system gives the expression of the Casimir force as \cite{50}
\bea
F_c(T,L)=-N[\frac{T}{T_o}]^{5/2}\frac{\pi^{2}\hbar^{2}}{mL^{3}}\   \ for \ T\le T_o
\eea
Here the exponent $5/2$ of $T/T_o$ is a consequence of the trapped geometry. 

\subsection{Conclusion}
             That vacuum fluctuation causes Casimir force is well known\cite{41}. Critical fluctuations also causes Casimir force\cite{43}. The Casimir force calculated here is neither due to vacuum fluctuations nor due to critical fluctuations. It is due to quantum fluctuations. These fluctuations are associated with the commutator algebra of position and momentum operator as well as with the commutator algebra of bosonic annihilation operator($\hat{a}_i$) and creation operator ($\hat{a}_i^{\dagger}$) such that $[\hat{a}_i,\hat{a}_j^{\dagger}]=\delta_{i,j}$, where $i,j$ represent the single particle energy states. For $T< T_o$ a macroscopic number of particles come down to the ground state. The quantum fluctuations die out due to the macroscopic occupation of particles in a single state. That is why the Casimir force dies out at $T< T_o$. For, $T\gg T_o$ the Bose-Einstein statistics becomes classical Maxwell-Boltzmann statistics and thermal fluctuations dominates over the quantum fluctuations. For this reason the Casimir force dies out at $T\gg T_o$.  For our system, the reduction of thermodynamic Casimir force with the power law will show the signature of the Bose-Einstein condensation. 
\appendix{}
\section{Zeta functions for negative integers}
Zeta functions for the negative integers are obtained from the analytic continuation technique \cite{51}. However, here we will present a simple technique. Our technique is as follows.

             Let us consider a function $f$ of a variable $t$, such that $f(t)=\frac{t}{e^t-1}$. We will expand this function in two different ways and will collect the coefficients of the powers of $t$. Equating the coefficients we will get the zeta functions for the negative integers.
\subsection{Expansion of $f(t)$}
\begin{eqnarray}
f(t)&=&\frac{t}{e^{t}-1}=te^{-t}(1-e^{-t})^{-1}\nonumber\\&=&t[e^{-t}+e^{-2t}+e^{-3t}+e^{-4t}+...]\nonumber\\=t&[&1-t+\frac{t^2}{2!}-\frac{t^3}{3!}+\frac{t^4}{4!}-....\nonumber\\+&1&-2t+\frac{2^2t^2}{2!}-\frac{2^3t^3}{3!}+\frac{2^4t^4}{4!}-....\nonumber\\+&1&-3t+\frac{3^2t^2}{2!}-\frac{3^3t^3}{3!}+\frac{3^4t^4}{4!}....\nonumber\\+&&..........]
\end{eqnarray}          
We see that the coefficient of $t^k$ of the above expansion(A1) goes as $(-1)^{k+1}\frac{\zeta(-k+1)}{(k-1)!}$ such that           
\bea
f(k)=\Sigma_{k=1}^{\infty}\frac{(-1)^{k-1}\zeta(-k+1)}{(k-1)!}t^k.
\eea
 
\subsection{Another expansion of $f(t)$}
\begin{eqnarray}
f(t)&=&\frac{t}{e^{t}-1}=\frac{t}{t+\frac{t^2}{2!}+\frac{t^3}{3!}+\frac{t^4}{4!}+...}=\frac{1}{1+\frac{t^1}{2!}+\frac{t^2}{3!}+\frac{t^3}{4!}+...}\nonumber\\&=&(1+\frac{t^1}{2!}+\frac{t^2}{3!}+\frac{t^3}{4!}+..)^{-1}\nonumber\\&=&1-(\frac{t^1}{2!}+\frac{t^2}{3!}+\frac{t^3}{4!}..)+\frac{(-1)(-2)}{2!}(\frac{t^1}{2!}+\frac{t^2}{3!}+\frac{t^3}{4!}+..)^2-\frac{(-1)(-2)(-3)}{3!}(\frac{t^1}{2!}+\frac{t^2}{3!}+\frac{t^3}{4!}+..)^3+....\nonumber\\&=&1-\frac{1}{2}t+(\frac{1}{4}-\frac{1}{6})t^2+(-\frac{1}{4!}+2\frac{1}{2!}\frac{1}{3!}-(\frac{1}{2!})^3)t^3+(-\frac{1}{5!}+\frac{1}{(3!)^2}+\frac{1}{4!}-3\frac{1}{(2!)^2}\frac{1}{3!}+\frac{1}{(2!)^4})t^4+...
\end{eqnarray}
The coefficients of $t^k$ the above expansion(A3) can be written in terms of Bernoulli's number$(B_k)$ such that
\bea
f(k)=\Sigma_{k=0}^{\infty}\frac{B_k}{k!}t^k.
\eea
From the above expansion(A3) we see that $B_0=1$, $B_1=-1/2$, $B_2=1/6$, $B_3=0$, $B_4=-1/30$, $B_5=0$ etc. 
\subsection{Relation between the Bernoulli's numbers and zeta functions}
Let us now equate the coefficients of $t^k$ of the two different expansions of $f(t)$.
Equating the coefficients we get
\bea
(-1)^{k+1}\frac{\zeta(-k+1)}{(k-1)!}=\frac{B_k}{k!} 
\eea
From the above eqn.(A5) we get
\bea
\zeta(-k)=(-1)^k\frac{B_{k+1}}{k+1}
\eea  
Now we see that $\zeta(0)=1+1+1+1+..=B_1/1=-1/2$, $\zeta(-1)=1+2+3+4+..=-B_2/2=-1/12$, $\zeta(-2)=1^2+2^2+3^2+4^2+..=B_3/3=0$, $\zeta(-3)=1^3+2^3+3^3+4^3+..=-B_4/4=1/120$, $\zeta(-4)=1^4+2^4+3^4+4^4+..=B_5/5=0$ etc.  Although our first expansion of $f(t)$ is not mathematically sound yet the results obtained for the $\zeta(-k)$ are the same as obtained from the analytic continuation technique\cite{51}.

\end{document}